# Significant Results from SUMER/SOHO*

B.N. Dwivedi

Department of Applied Physics, Institute of Technology,
Banaras Hindu University, Varanasi-221005, India
email: bholadwivedi@yahoo.com

**Abstract.** We briefly outline recent observations by solar spacecraft such as Yohkoh, SOHO, TRACE, and RHESSI, which have revolutionized what we know and don't know about the Sun. We then present some significant results, mainly from SUMER/SOHO but also complimentary from the other SOHO's experiments, such as CDS, EIT, UVCS, and LASCO. In particular, we present density-temperature structures, explosive events, velocity anisotropy, wave activity, coronal holes and the solar wind etc. These results have provided valuable clues to a better understanding of the two of the SOHO's principal scientific objectives namely, how the Sun's magnetic energy heats its million-degree corona, and feeds the solar wind.

## 1. Introduction

Spacecraft, built by the U.S. and Soviet space agencies in the 1960's and 1970's, dedicated to solar observations, added much to our knowledge of the Sun's atmosphere; notably the manned NASA Skylab mission of 1973-1974. Ultraviolet and X-ray telescopes on board gave the first high-resolution images of the chromosphere and corona and the intermediate transition region. Images of active regions revealed a complex of loops which varied greatly over their lifetimes, while ultraviolet images of the quiet Sun showed that the transition region and chromosphere followed the 'network' character previously known from the Ca II K-line images. Figure 1 illustrates the chromosphere in the emission line of neutral helium at 58.43 nm formed at about 20,000 K. More recently, the spatial resolution of spacecraft instruments has steadily improved, nearly equal to what can be achieved with ground-based solar telescopes (Dwivedi and Phillips, 2003). Each major solar spacecraft since Skylab has offered a distinct improvement in resolution. From 1991 to late 2001, the X-ray telescope on the Japanese Yohkoh spacecraft has routinely imaged the Sun's corona, tracking the evolution of loops and other features through its 11-year cycle of solar activity (cf., Figure 2). For comprehensive reviews on topics from the Sun's interior to its exterior, including the solar wind and the solar observing facilities, the reader is referred to Dwivedi (2003).

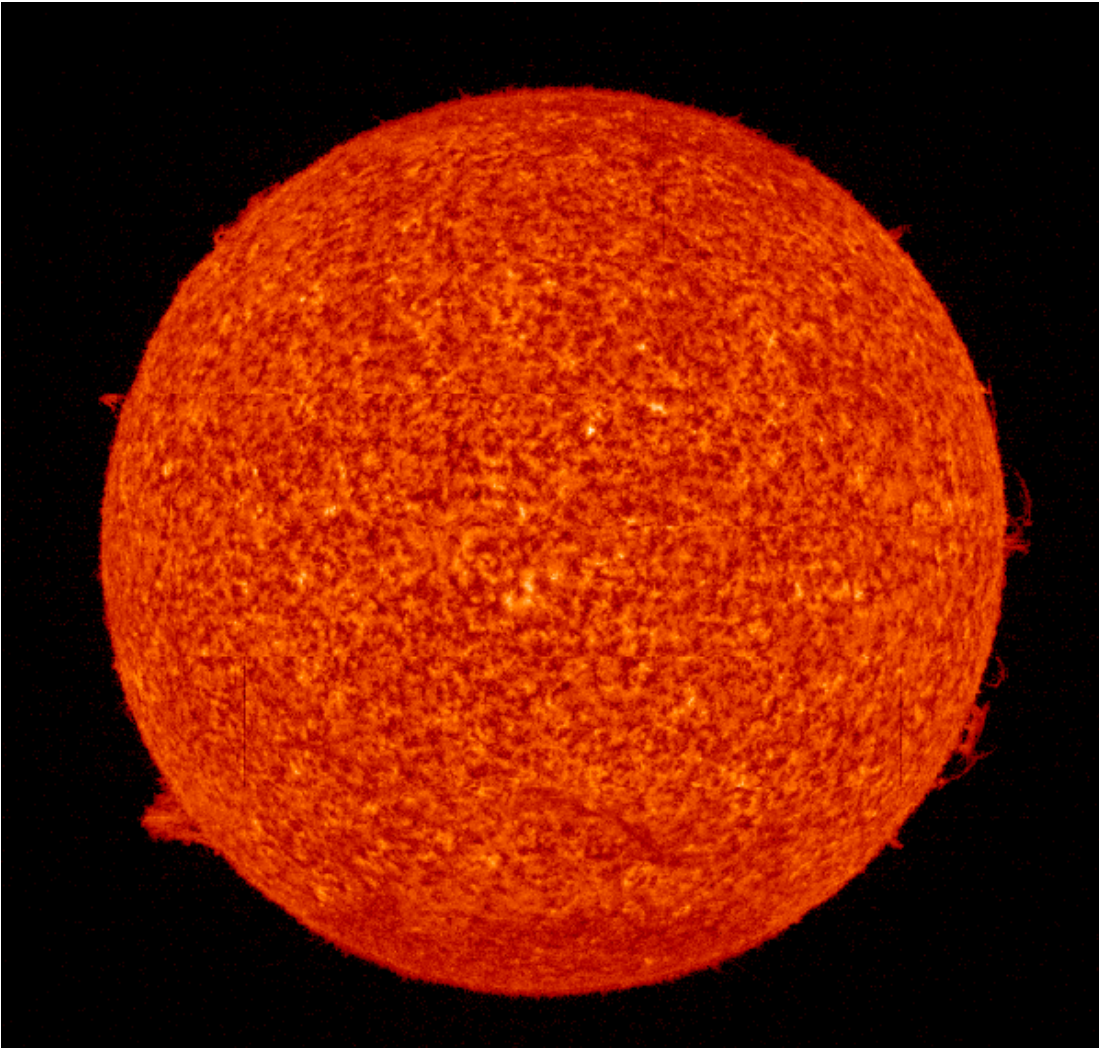

**Figure 1.** The Sun's image in the chromospheric emission line of neutral helium at 58.43 nm wavelength taken by the SUMER spectrograph on the SOHO spacecraft on 4 March 1996. (Credit : SUMER/SOHO, ESA-NASA).

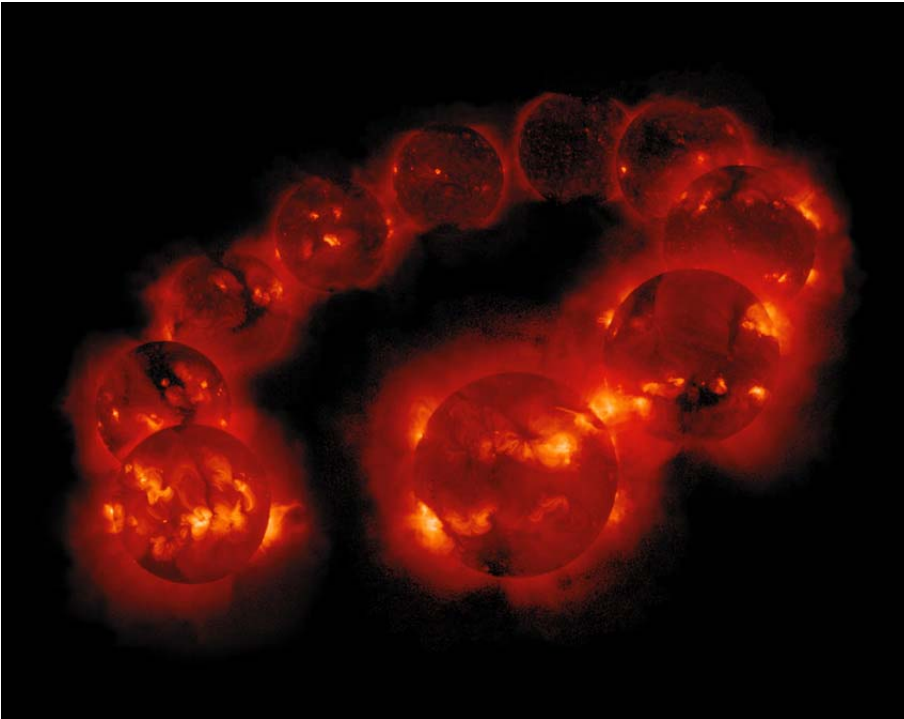

**Figure 2.** This image shows the solar corona in X-rays as observed by Yohkoh's Soft X-ray Telescope (SXT). The series of images are from 8 January 1992 (lower left) to 25 July 1999 (lower right). The X-ray images show much more striking changes in patterns and intensities, driven by the changes in the magnetic field. The hot corona extends hundreds of thousands of kilometers above the Sun's surface and is shaped into complex forms by magnetic fields (Credit : SXT/Yohkoh).

The ESA/NASA Solar and Heliospheric Observatory (SOHO) was launched on December 2, 1995 into an orbit about the inner Lagrangian (L1) point situated some 1.5 x $10^6$ km from the Earth on the sunward side. Its twelve instruments, therefore, get an uninterrupted view of the Sun (Dwivedi and Mohan, 1997). These instruments are: GOLF (Global Oscillations at Low Frequency), VIRGO (Variability of solar Irradiance and Gravity Oscillations), SOI/MDI (Solar Oscillations Investigation/Michelson Doppler Imager), SUMER (Solar Ultraviolet Measurements of Emitted Radiation), CDS (Coronal Diagnostic Spectrometer), EIT (Extreme-ultraviolet Imaging Telescope), UVCS (Ultraviolet Coronagraph Spectrometer), LASCO (Large Angle Spectroscopic Coronagraph), SWAN (Solar Wind Anisotropies), CELIAS (Charge, Element and Isotope Analysis system), COSTEP (Comprehensive Suprathermal and Energetic Particle analyser), and ERNE (Energetic and Relativistic Nuclei and Electron experiment). The SOHO's scientific objectives include: what are the structure and dynamics of the solar interior ?, why does the corona exist and how is it heated ?, and where and how the solar wind accelerated ? Apart from a period in 1998 when the spacecraft was temporarily out of contact, it has been in continuous operation since launch. There are several imaging instruments, sensitive from visible-light wavelengths to the extreme-ultraviolet. The EIT, for instance, uses normal incidence optics to get full-Sun images several times a day in the wavelengths of lines emitted by the coronal ions Fe IX/Fe X, Fe XII, Fe XV (cf., Figure 3) as well as the chromospheric He II 30.4 nm line. The CDS and the SUMER are two spectrometers operating in the extreme-ultraviolet region, capable of getting temperatures, densities and other information from spectral line ratios. The UVCS has been making spectroscopic observations of the extended corona from 1.25 to 10 solar radii from the Sun's center, determining empirical values for densities, velocity distributions and outflow velocities of hydrogen, electrons, and several minor ions.

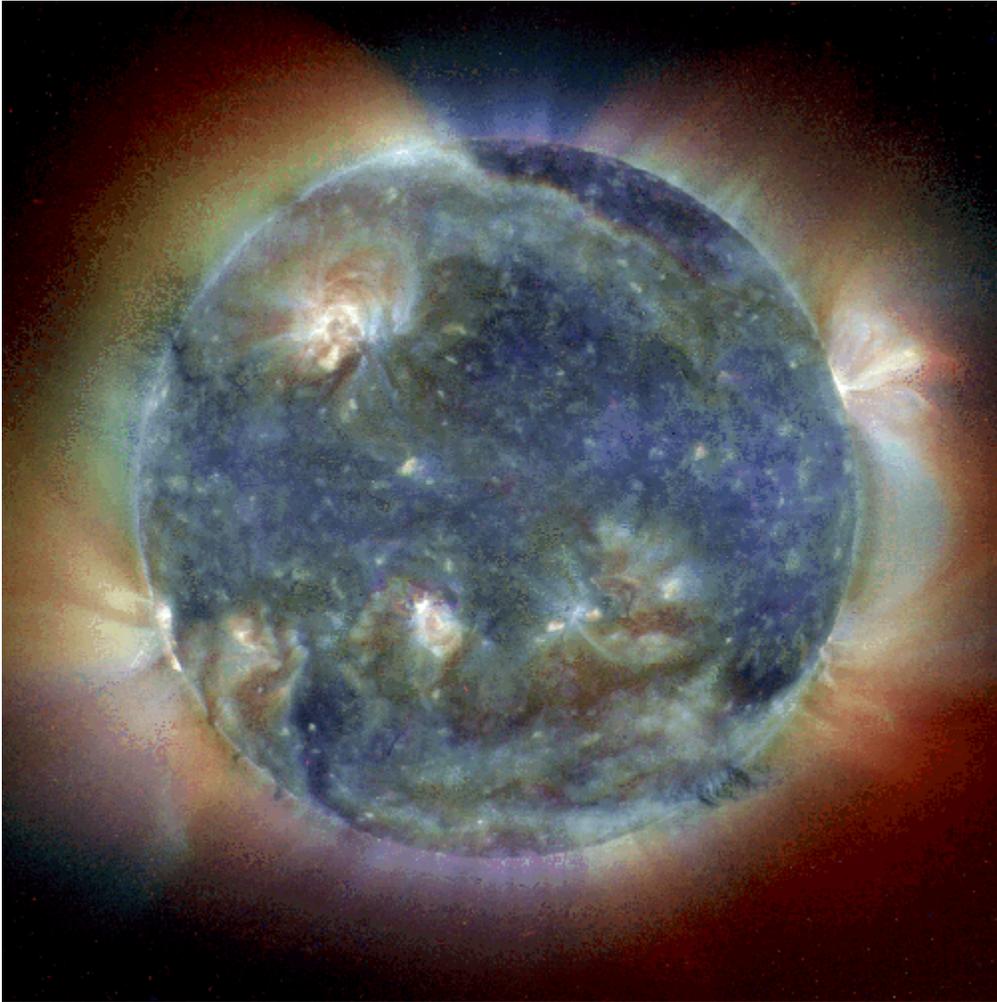

**Figure 3.** This tantalizing picture is a false colour composite of three images, all in extreme-ultraviolet light. Each individual image highlights a different temperature regime in the Sun's hot atmosphere and was assigned a different colour : red at $2 \times 10^6$ K, green at $1.5 \times 10^6$ K and blue at $10^6$ K. The combined image shows bright active regions strewn across the solar disk, which would otherwise appear as dark groups of sunspots in visible light images, along with some magnificent plasma loops and an immense prominence at the right hand solar limb. (Credit : EIT/SOHO, ESA-NASA).

The Transition Region and Coronal Explorer (TRACE) satellite went into a polar orbit around Earth in 1998. The spatial resolution is of order 1 arcsecond (725 km), and there are wavelength bands covering the Fe IX, Fe XII, and Fe XV lines as well as the Ly-alpha line at 121.6 nm. Its ultraviolet telescope has obtained images containing tremendous amount of small and varying features, for instance, active region loops are revealed to be only a few hundred kilometers wide, almost thread-like compared with their huge lengths. Their constant flickering and jouncing hint at the corona's heating mechanism. There is a clear relation of these loops and the larger arches of the general corona to the magnetic field measured in the photospheric layer. The crucial role of this magnetic field has only been realized in the past decade. The fields dictate the transport of energy between the surface of the Sun and the corona. The loops, arches and holes appear to trace out the Sun's magnetic field (cf., Figure 4a,b). The latest in the fleet of spacecraft dedicated to viewing the Sun is the Reuven Ramaty High Energy Solar Spectroscopic Imager (RHESSI), launched in 2002, which is providing images and spectra in hard X-rays (wavelengths less than about 4 nm). RHESSI's observations of tiny microflares (cf., Figure 5) may provide clues to the coronal heating mechanism.

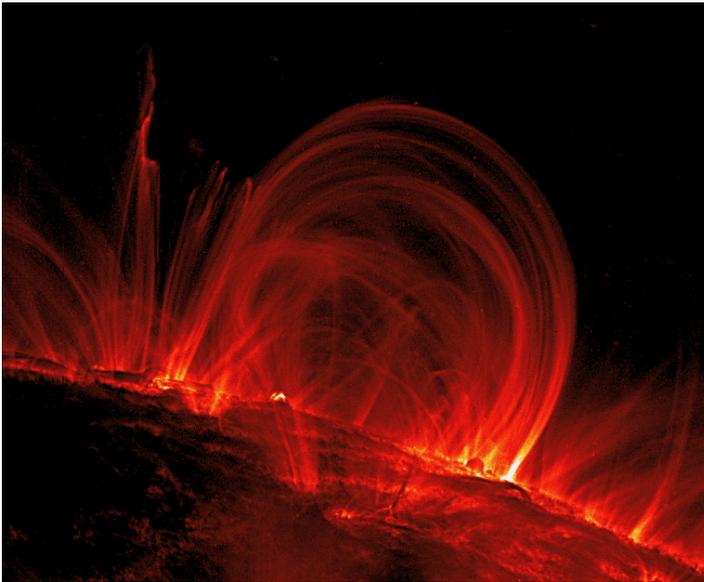

**Figure 4a.** Coronal loops, observed in the ultraviolet light (Fe IX 171 Å) by the TRACE spacecraft on 6 November 1999, extending 120,000 km off the Sun's surface (Credit : TRACE/NASA).

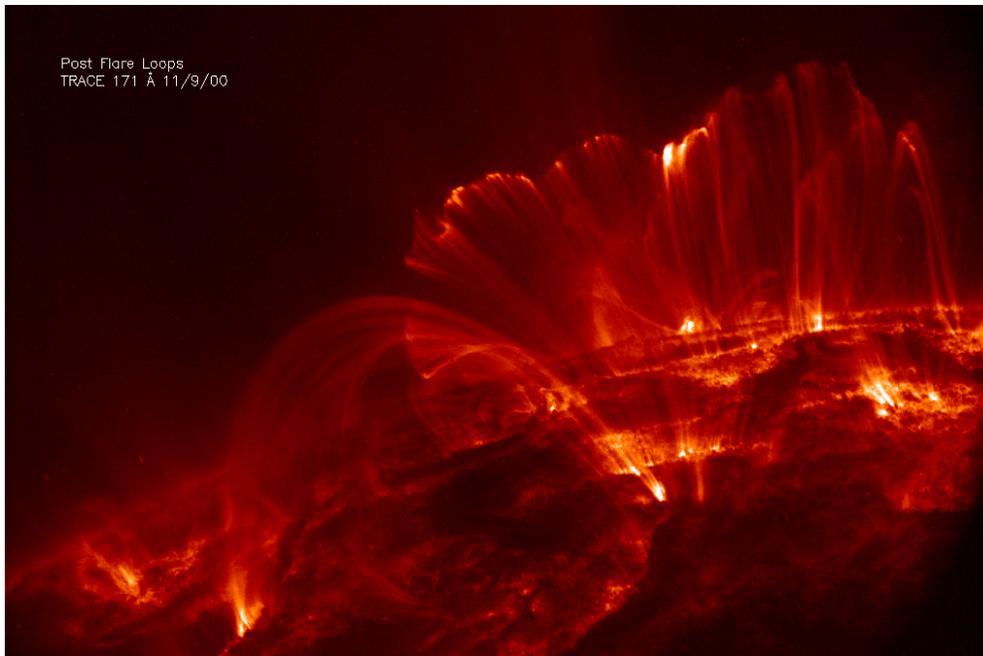

**Figure 4b.** Post flare loops, observed in the ultraviolet light (Fe IX 171 Å) by the TRACE spacecraft on 9 November 2000. (Credit : TRACE/NASA).

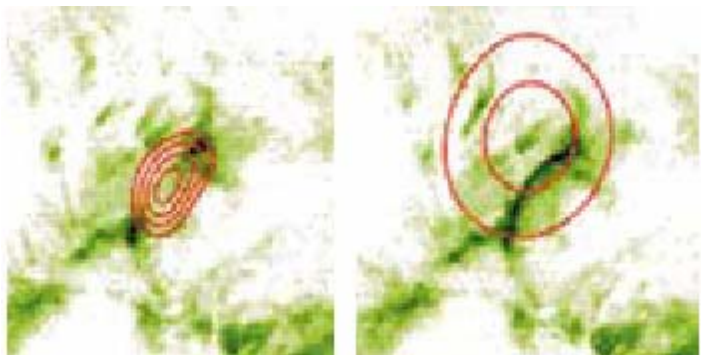

**Figure 5.** X-ray image taken by the RHESSI spacecraft outlines the progression of a microflare on 6 May 2002. The flare peaked (left), then six minutes later (right) began to form loops over the original flare site. (Credit : S. Krucker).

**The SUMER Spectrograph:** The Sun's extreme-ultraviolet (EUV) and UV spectra, in the wavelength range from 465 to 1610 Å which is the spectral range of the SUMER spectrograph, provide unique opportunities for probing the solar atmosphere from the chromosphere to the corona (cf., Figure 6). Emission-line intensities and their ratios and line shapes are used to obtain diagnostic information on plasmas in the temperature range from $10^4$ to more than $10^6$ K, allowing the derivation of characteristics such as temperature, density, abundance, turbulence, and flows. A full description of SUMER spectrograph and its performance are available (Wilhelm et al.,1995; Wilhelm et al.,1997; Lemaire et al., 1997; Wilhelm et al., 2004a). Briefly, the instrument is an ultraviolet telescope and spectrometer with a wavelength resolution element of 42–44 mÅ over the range 800−1610 Å (in first order). Along the north-south directed slit, the spatial resolution is close to 1 arcsec (≈715 km on the Sun). The spectral resolution, which is twice as high in second order, can further be improved for relative (and under certain conditions, for absolute) measurements to fractions of a pixel, allowing the measurement of Doppler shifts corresponding to plasma bulk velocities of about 1 km/sec along the line of sight. There now exist a vast literature over the last eight years or so on the results from SUMER observations. Obviously, we cannot review this literature in this article. Instead, we present some significant results from SUMER on solar plasma that have added new dimensions to a better understanding of the solar mysteries, especially coronal heating and the solar wind acceleration.

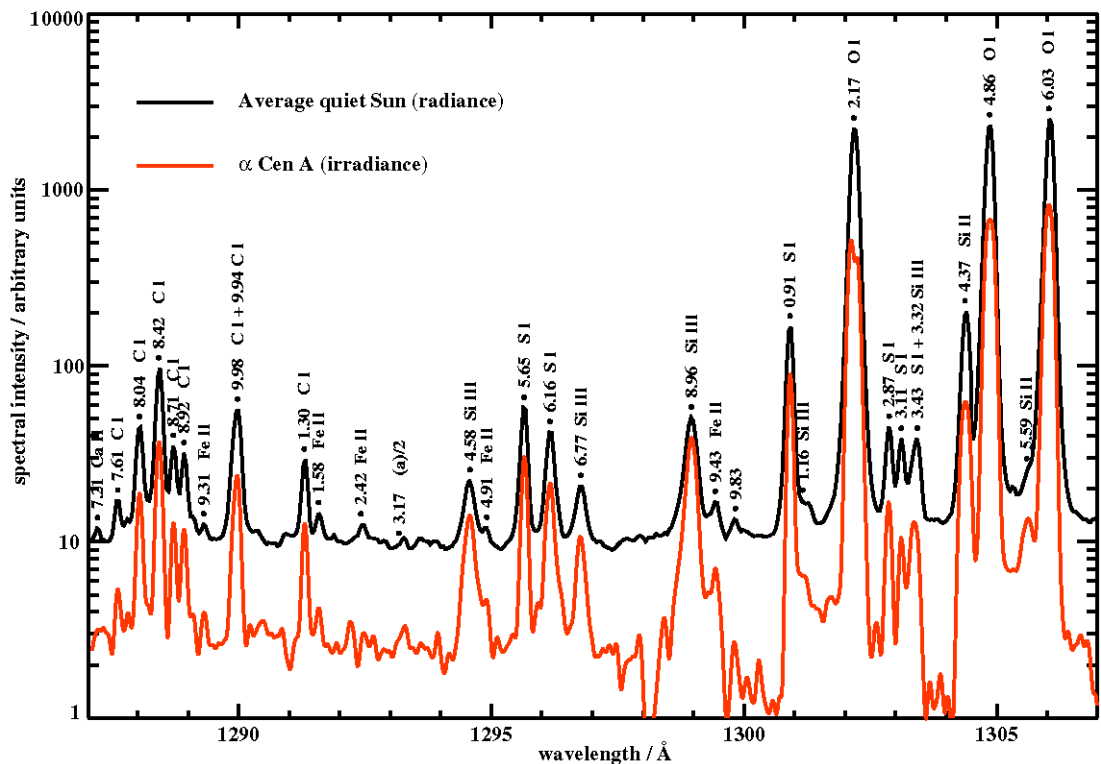

**Figure 6.** The SUMER spectral atlas is the best-ever analysis of the ultraviolet light from the Sun, spanning wavelengths from 670 to 1609 Ångstrom (67 to 160.9 nanometers), and identifies some 1100 distinct emission lines, of which more than 150 had not been recorded or identified before SOHO. (Credit: SUMER/SOHO).

We can directly measure physical parameters such as electron density, temperature, flow speeds, etc. in the corona from emission line diagnostics. However, we cannot directly measure magnetic field strength, resistivity, viscosity, turbulence, waves, etc. New powerful tools of coronal seismology have enabled the detection of MHD waves by TRACE and EIT, spectroscopic measurements of line-widths by SUMER and CDS, ion and electron temperature anisotropy measurements with UVCS, and microflares by RHESSI.

## 2. Density – Temperature Structures

Without a knowledge of the densities, temperatures and elemental abundances of space plasmas, almost nothing can be said regarding the generation and transport of mass, momentum and energy. Thus, since early in the era of space-borne spectroscopy we have faced the task of inferring plasma temperatures, densities and elemental abundances for hot solar and other astrophysical plasmas from optically thin emission-line spectra (Dwivedi, 1994; Mason and Monsignori-Fossi, 1994; Dwivedi et al., 2003; Wilhelm et al., 2004a). A fundamental property of hot solar plasmas is their inhomogeneity. The emergent intensities of emission lines from optically thin plasmas are determined by integrals along the line of sight (LOS) through the plasma. Spectroscopic diagnostics of the temperature and density structures of hot optically thin plasmas using emission-line intensities is usually described in two ways. The simplest approach, the line-ratio diagnostics, uses an observed line-intesity ratio to determine density or temperature from theoretical density or temperature-sensitive line-ratio curves, based on an atomic model and taking account of physical processes for the line formation. The line-ratio method is stable, leading to well-defined values of $N_e$ or $T_e$ , but in realistic cases of inhomogeneous plasmas, these are hard to interpret. The more general differential emission measure (DEM) method recognizes that observed plasmas are better described by distributions of temperature or density along the LOS, and poses the problem in inverse form. It is well known that the DEM function is the solution to the inverse problem, which is function of $N_e$ or $T_e$ or both. Derivation of DEM functions, while more generally acceptable, is unstable to noise and errors in spectral and atomic data. The exact relationship between the two approaches has been investigated (Brown et al., 1991 ; McIntosh et al., 1998).

Line shifts and broadenings give information about the dynamic nature of the solar and stellar atmospheres. The transition region spectra from the solar atmosphere are characterized by broadened line profiles. The nature of this excess broadening puts constraints on possible heating processes. Systematic redshifts in transition region lines have been observed in both solar and stellar spectra of late-type stars. On the Sun, outflows of coronal material have been correlated with coronal holes. The excess broadening of coronal lines above the limb provides information on wave propagation in the solar wind (cf., Figure 7; Kohl et al., 1998)

# O VI $\lambda\lambda$ 1032, 1037 line widths

## SOLAR DISK   (SUMER/SOHO)

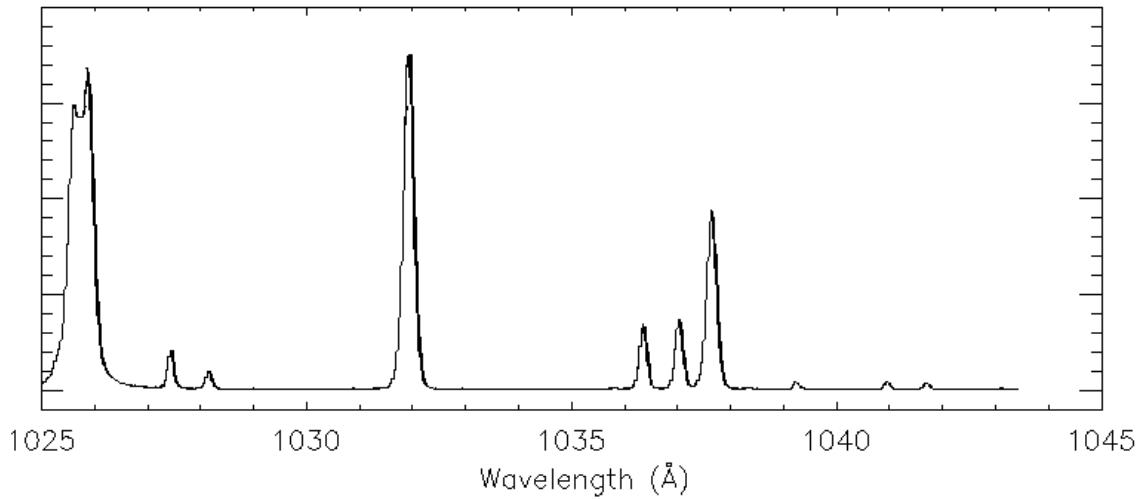

## N. Pole, 2.1 $R_\odot$   (UVCS/SOHO)

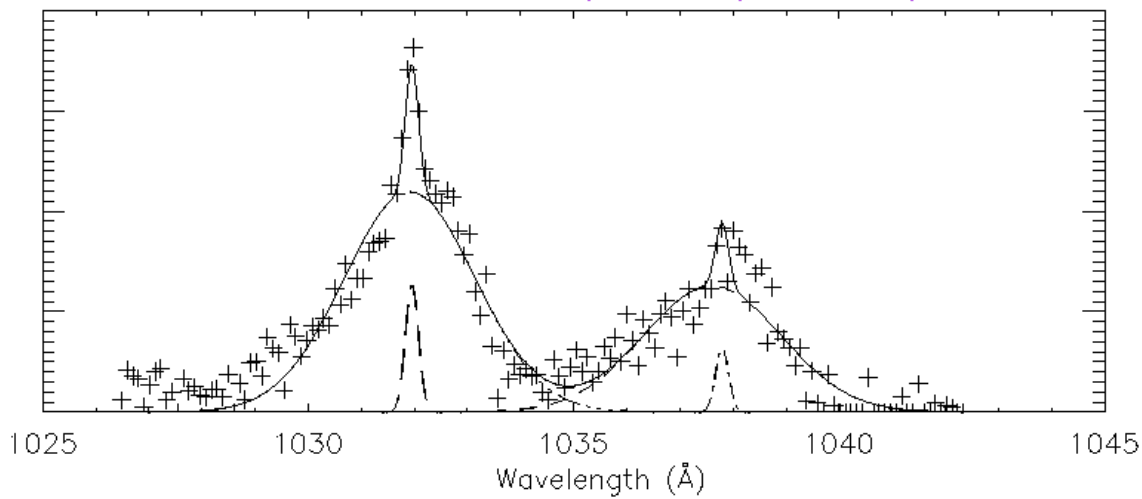

**Figure 7.** Striking difference in the width of line profiles observed on disk and in a polar coronal hole --- Discovery of the large velocity anisotropy. Solar wind acceleration by ion-cyclotron resonance (Kohl et al. 1998).

### 3. Explosive Events

The universe abounds with explosive energy release that may heat plasma to millions of degrees and accelerate particles to relativistic velocities. Such occurrences are not uncommon on our own star too. Examples include solar flares, coronal mass ejections, chromospheric and coronal microflares, etc. In many cases, the magnetic field seems to be the only source of energy available to power these cosmic explosions. While it is well established that the Sun has a large reservoir of magnetic energy, the reason for its release is still debated. Explosive events were first seen in the ultraviolet spectra obtained with the High Resolution Telescope and Spectrograph (HRTS) flown on several rocket flights and Spacelab 2 (Dere et al., 1991). The SUMER spectrograph made it possible to observe the chromospheric network continuously over an extended period and to discern the spatial structure of the flows associated with these explosive events. The observation that explosive events are bi-directional jets provides new evidence that they result from magnetic reconnection above the solar surface.

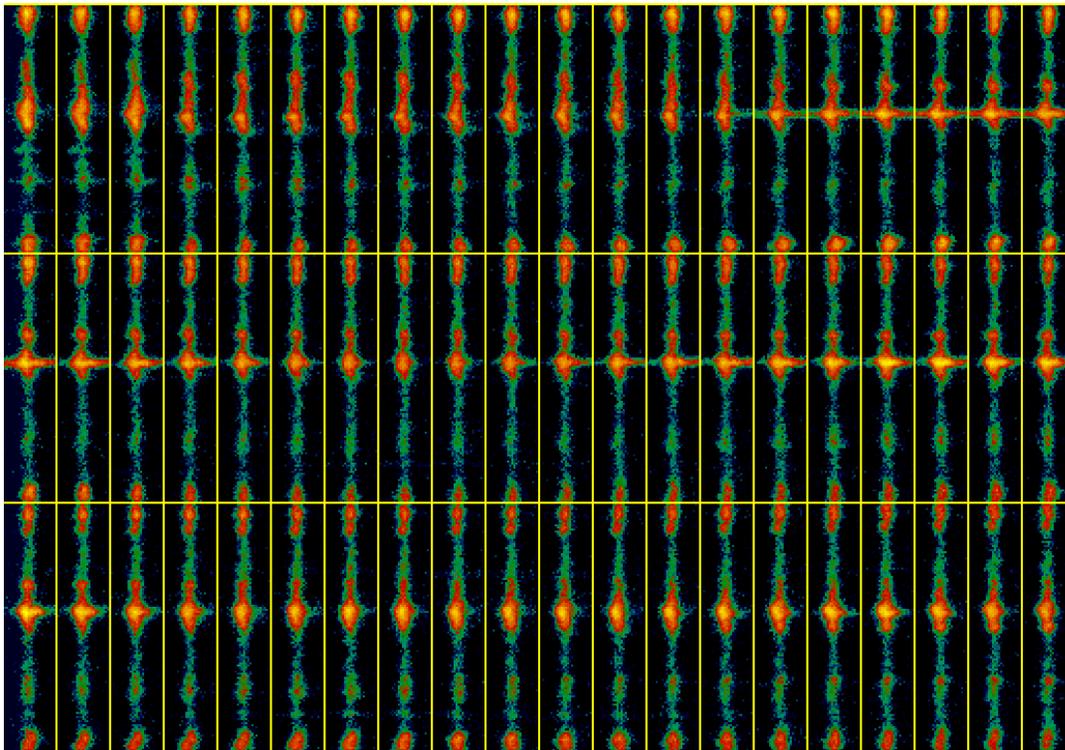

**Figure 8.** Explosive events seen in the Si IV 1393 Å line. The SUMER spectrograph slit shown in each section has a projected length of 84000 km on the Sun, and a width of 700 km. The exposure time was 10 sec each. The Doppler shift near bright portions (i.e., at chromospheric network crossings of the slit) corresponds to plasma motions with line-of-sight velocities ±150 km/sec. (Credit: SUMER/SOHO, ESA-NASA).

Interpreting the evolution of the jets in Si IV 1393 Å line profile, Innes et al. (1997) have shown that explosive events have the bi-directional jets ejected from small sites above the solar surface. Figure 8 shows the explosive events seen in the Si IV 1393 Å line. The SUMER spectrograph slit shown in each section has a projected length of 84000 km on the Sun, and a width of 700 km. The exposure time was 10 sec each. The Doppler shift near bright portions corresponds to plasma motions with line of sight velocities ±150 km/sec. The structure of these plasma jets evolves in the manner predicted by theoretical models of magnetic reconnection. This lends support to the view that magnetic reconnection is one of the fundamental processes for accelerating plasma on the Sun. Such observations seem to provide the best evidence to date for the existence of bi-directional outflow jets which is a fundamental part of the standard magnetic reconnection model. This provides a new clue to our understanding of how the Sun's magnetic energy feeds its million degree hot corona, and the solar wind.

## 4. Wave Activity

Hassler et al. (1990) carried out rocket-borne experiment to observe off-limb line width profile of Mg X 609 and 625 Å and reported the increasing line width with altitude (up to 70,000 km). This observation provided the signature of outward propagating undamped Alfvén waves. SUMER and CDS spectrometers recorded several line profiles and found the broadening of emission lines. These results are consistent with outward propagation of undamped Alfvén waves travelling through regions of decreasing density (Erdélyi et al., 1998). Harrison et al. (2002) reported the narrowing of the Mg X 62.50 nm line with height in the quiet near-equatorial solar corona, and concluded that this narrowing is likely evidence of dissipation of Alfvén waves in closed field-line regions. Similarly, a significant change in slope of the line width as a function of height was seen in polar coronal holes by O'Shea et al. (2003) at an altitude of ≈65 Mm. These results obtained with the CDS, if confirmed, could be crucial in understanding the coronal heating mechanisms. Due to the broad instrumental profile, the CDS instrument can only study line-width variations and cannot provide measurements of the line width itself, and, hence, of the effective ion temperature. Since the latter quantity is critical in constraining theoretical models of coronal heating and solar-wind acceleration, for instance, through the dissipation of high-frequency waves generated by chromospheric reconnection, Wilhelm et al. (2004b) studied the problem further by analysing data recorded with the SUMER spectrograph in the Mg X doublet together with other neighbouring lines in both the quiet equatorial corona and in a polar coronal hole. Due to the high spectral resolution of SUMER, they were able to obtain profiles of both Mg X emission lines and measure their widths and variations as a function of height. Their work showed that line widths of both components of the Mg X doublet measured by SUMER monotonically increase in the low corona in equatorial regions in altitude ranges for which scattered radiation from the disk does not play a major rôle. They did not find any evidence for a narrowing of the emission lines above 50 Mm. The same statement applies for a coronal hole, but they could not exclude the possibility of a constant width above 80 Mm (cf., Figure 9).

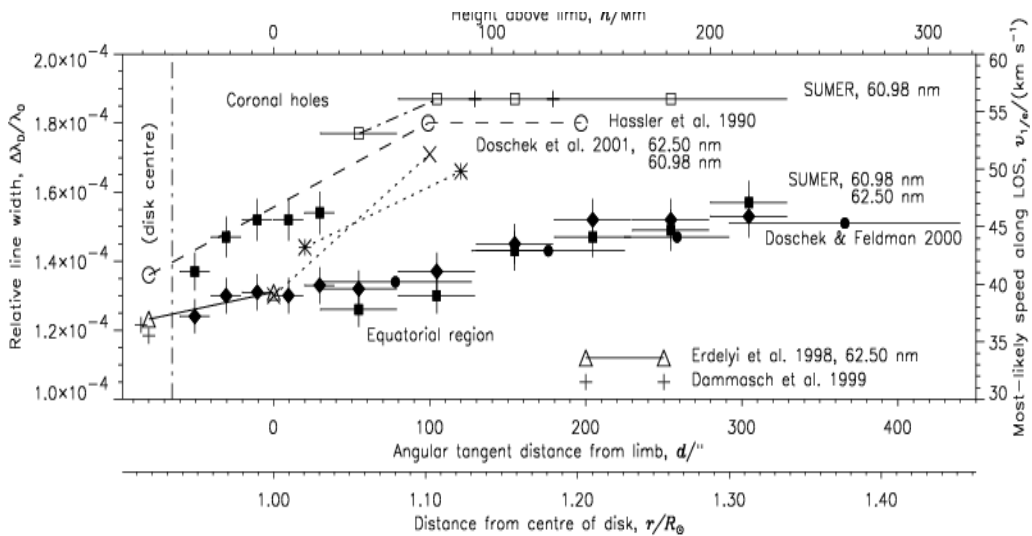

**Figure 9.** Relative line-width variations as a function of radial distance from the Sun. Literature data and results obtained in this work (annotated "SUMER") are compiled for the Mg X doublet in equatorial and coronal-hole regions. A classification of the Hassler et al. observations, of which only three typical values are shown, is not defined in their 1990 paper. An approximate scale of 715 km/ ″has been used for both the SOHO observations and those from the Earth. Integration intervals and the SUMER uncertainty margins are marked by horizontal and vertical bars. Related data points are in some cases connected by lines of various styles. They are meant to improve the orientation of the reader, but not as physical interpolations, in particular, for those points representative of centre-of-disk values displayed here near -80 ″. (From Wilhelm et al. 2004b).

While there have been reports of emission-line broadening with altitude, using SUMER, the CDS observations presented by Harrison et al. (2002), already noted, appeared to show emission-line narrowing. In order to resolve the apparent discrepancies, a joint CDS/SUMER observational sequence was successfully executed during the SOHO/MEDOC campaign in November and December 2003. The joint measurements were performed near the east limb (cf., Wilhelm et al. 2005). The pointing locations of the spectrometers are shown in Figs. 10a and 10b superimposed on He II and Fe XII solar images taken by EIT (Wilhelm et al. 2005). In the western corona, SUMER made additional exposures with a similar observational sequence from 13:02 to 15:06 UTC. Since the corona was very hot there, the slit positions are shown together with the Fe XII and Fe XV windows of EIT in Figs. 10c and 10d.

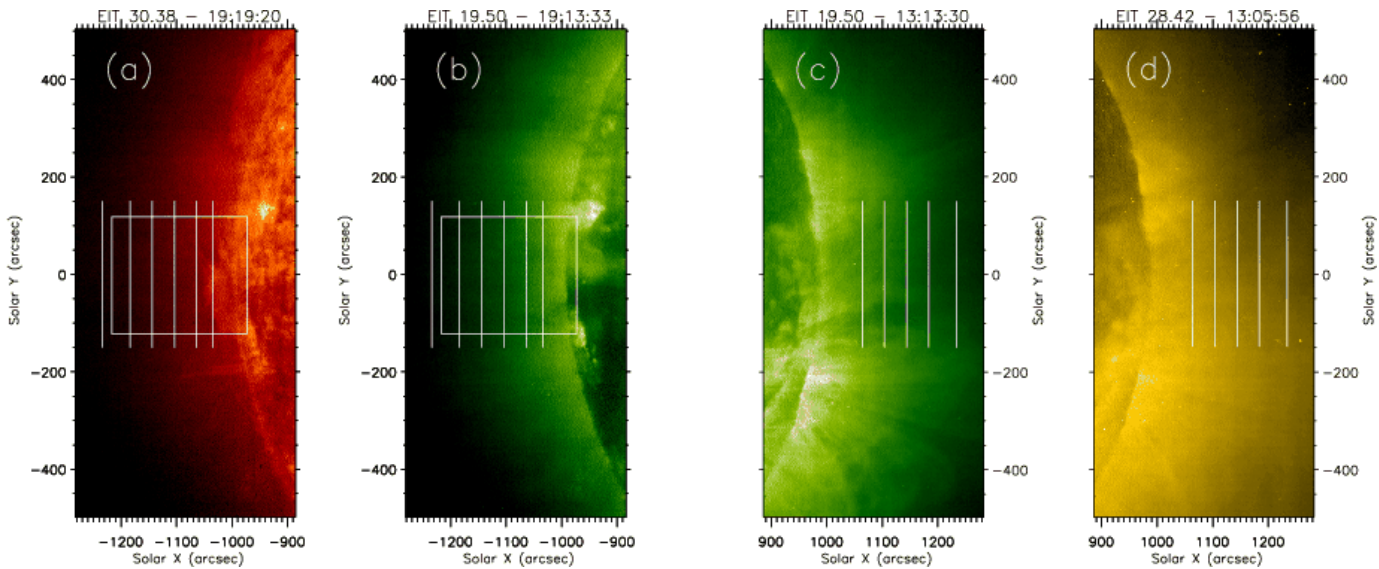

**Figure 10.** Positions of the CDS FOVs (rectangles) and the SUMER slit pointing locations in relation to He II, Fe XII and Fe XV solar images of 4 December 2003 (courtesy of the EIT consortium). Joint observations were obtained in the eastern corona. At low altitudes, a prominence caused a slight disturbance there. (a) He II spectral window; and (b) Fe XII window near the east limb; (c) Fe XII window; and (d) Fe XV window near the west limb. (From Wilhelm et al. 2005).

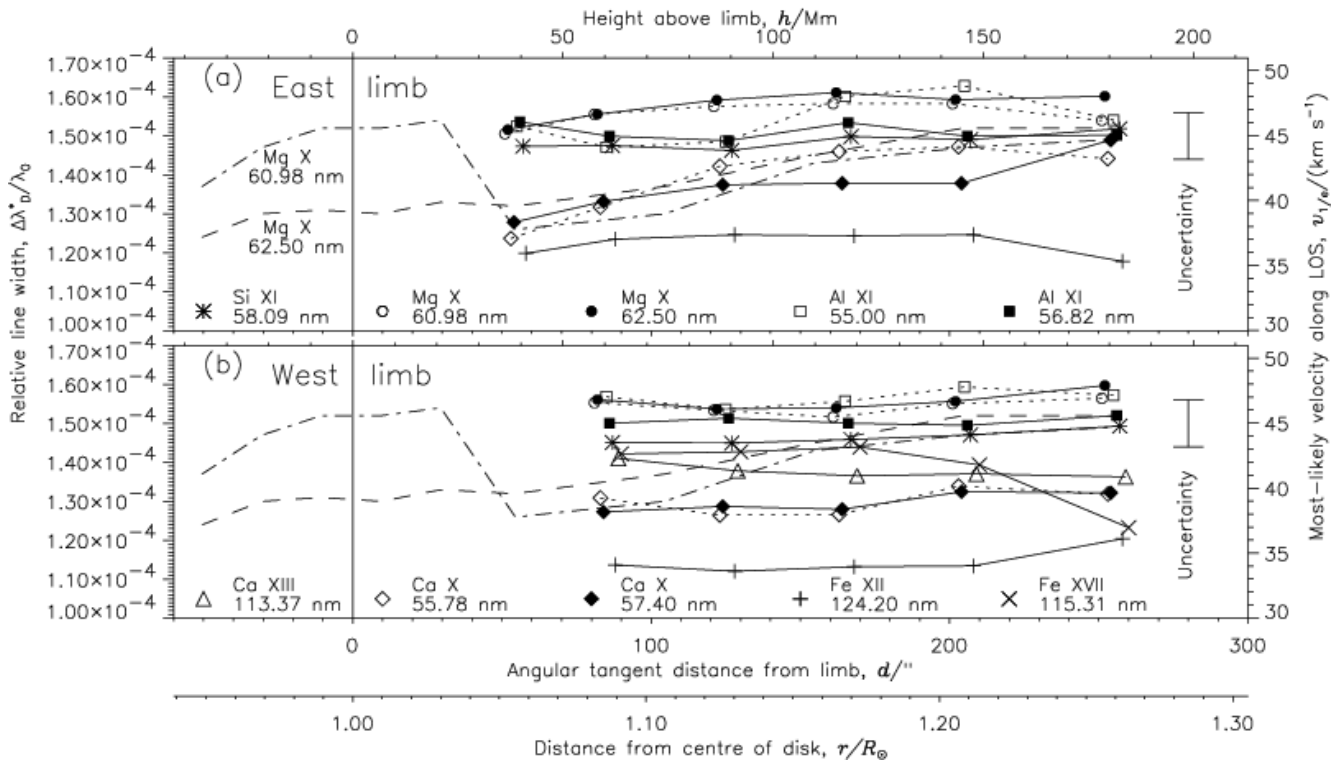

**Figure 11.** Summary of the SUMER line-width measurements in terms of the relative line width, $\Delta\lambda_D{}^{\dagger}/\lambda_0$, of the various ions, as well as their most-likely velocity along the LOS, $v_{1/e}$, as a function of distance from the limb. (a) Measurements above the east limb; (b) above west limb. The dashed and dash-dotted lines indicate the Mg X observations of SUMER in November 1996. The increase of the width of the 60.98 nm line near the limb is caused by blends of O III and O IV transition-region lines (cf. Wilhelm et al. 2004). Near the centre of the solar disk a value of $\Delta\lambda_D{}^{\dagger}/\lambda_0 \approx 1.2 \times 10^{-4}$ was observed for Mg X (Erdélyi et al. 1998; Dammasch et al. 1999). At each slit position, an altitude range of $\approx$ 10 ″ was covered. In the interest of clarity, we spread the plot symbols of the various spectral lines over this range. (From Wilhelm et al. 2005).

The relative widths of the spectral lines observed by SUMER are shown in Figure 11, separately for the eastern and western FOVs. In cases for which more than one observation was available, mean values have been shown. The range of $v_{1/e}$ extends from 35 km s$^{-1}$ to 49 km s$^{-1}$ in the relatively quiet eastern corona, and from 33 km s$^{-1}$ to 48 km s$^{-1}$ in the active western corona. No significant differences could be noticed within the uncertainty margins. The Mg X and Ca X lines show slight increases with height in the east, but are rather constant in the west. The Fe XII line is a little narrower in the west. Of particular interest was that the lines of Ca XIII and Fe XVIII were seen in the west, where they could be compared with the Ca X and Fe XII lines along the same LOS.

In the relatively quiet equatorial corona above a small prominence, Wilhelm et al. (2005) found none or very slight increase of the line widths of coronal emission lines with altitude from measurements both with CDS and SUMER. Taking the combined uncertainty margins into account, the relative variations for Mg X were considered to be consistent, although the absolute widths could not be compared given the different instrumental spectral transfer functions. The SUMER observations indicated even less line-width variations with height in the more active corona. Singh et al. (2003) found height variations of line profiles in the visible light that depended on the formation temperatures of the lines. So, the solar conditions appear to have a direct influence on the line-width variations with height. Whether this dependence can account for the past apparent discrepancies cannot unambiguously be decided with the information available. More observations for different coronal activity levels are needed for this task. However, this joint study concluded that CDS and SUMER relative line-width measurements did not lead to inconsistencies if the same solar region is under study.

## 5. Coronal Holes and the Solar Wind

Observations from the Skylab firmly established that the high-speed solar wind originates in coronal holes which are well-defined regions of strongly-reduced ultraviolet and X-ray emissions (Zirker 1977). More recent data from Ulysses show the importance of the polar coronal holes, particularly at times near the solar minimum, when dipole field dominates the magnetic field configuration of the Sun. The mechanism for accelerating the wind to the high values observed, of the order of 800 km sec$^{-1}$, is not yet fully understood. The Parker model is based on a thermally-driven wind. To reach such high velocities, temperatures of the order 3 to 4 MK would be required near the base of the corona. However, other processes are available for acceleration of the wind, for example, the direct transfer of momentum from MHD waves, with or without dissipation. This process results from the decrease of momentum of the waves as they enter less dense regions, coupled with the need to conserve momentum of the total system. If this transfer predominates, it may not be necessary to invoke very high coronal temperatures at the base of the corona.

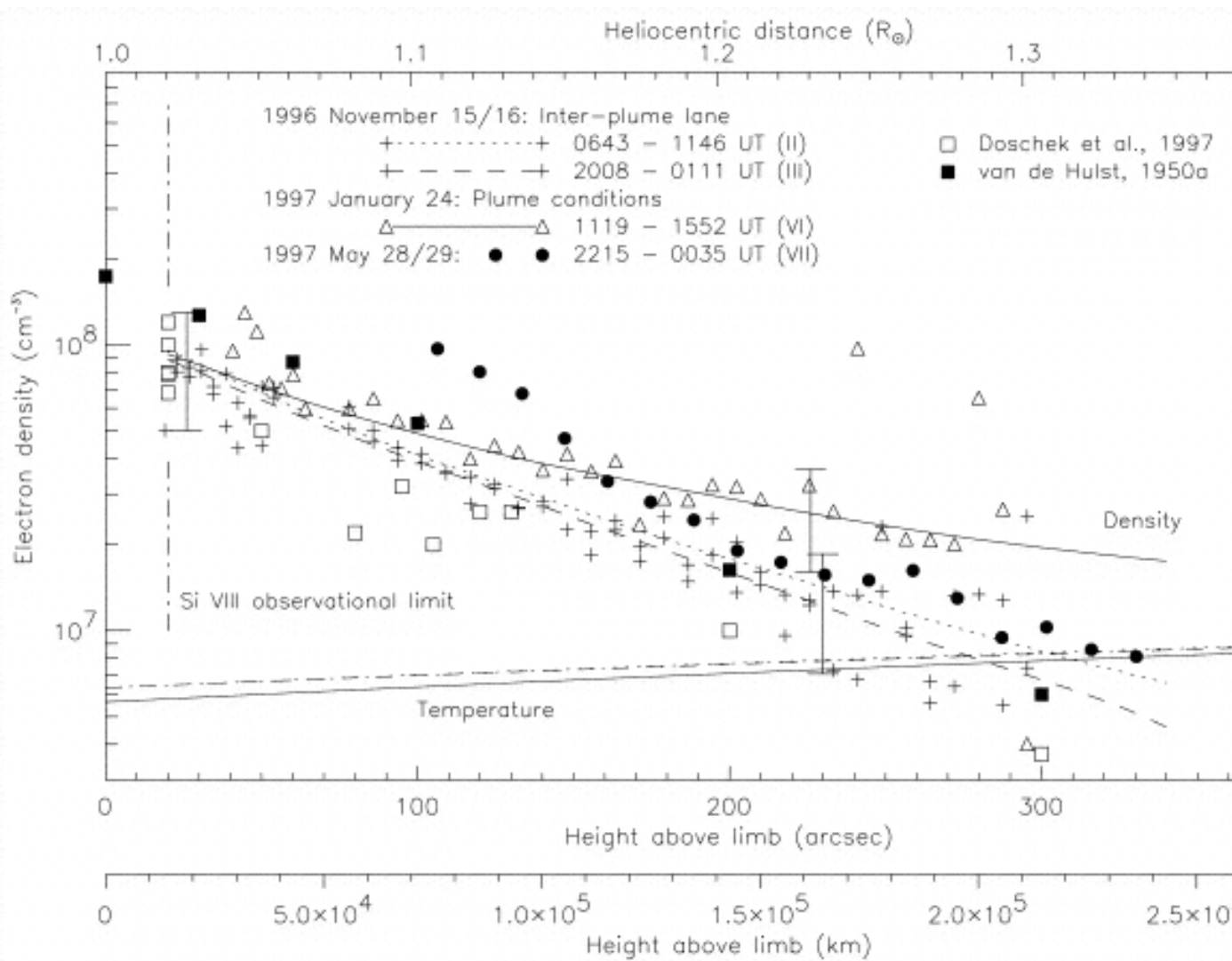

**Figure 12a.** Electron densities derived from Si VIII line ratios and a comparison with data in the literature. The hydrostatic temperature, $T_S$, used for the fits of the line Si VIII (1445 Å) is plotted in the lower portion of the diagram, with a scale on the right-hand side. The points labeled "1997 May 28/29" are obtained from the ratios observed west of the polar plume assembly in a very dark region of the corona. The error bars indicate a density variation resulting from a ±30% uncertainty in the line ratio determination. Note that there is a small (3%) seasonal variation between the angular and the spatial scales for different data sets. (From Wilhelm et al. 1998).

Prior to the SOHO mission, there was very little information available on the density and temperature structure in coronal holes. Data from Skylab was limited, due to the very low intensities in holes and poor spectral resolution, leading to many line blends. High-resolution ultraviolet observations from instruments on SOHO spacecraft provided the opportunity to infer the density and temperature profile in coronal holes (cf., Figure 12a, Wilhelm et al. 1998). Comparing the electron temperatures with ion temperatures, it was concluded that ions are extremely hot and the electrons are relatively cool. Using the CDS and SUMER instruments on the SOHO spacecraft, electron temperatures were measured as a function of height above the limb in a coronal hole. Observations of two lines from the same ion, O VI 1032 Å from SUMER and O VI 173 Å from CDS, were made to determine temperature gradient in a coronal hole (David et al. 1998). This way temperature of around 0.8 MK close to the limb was deduced, rising to a maximum of less than 1 MK at 1.15 $R_\odot$ , then falling to around 0.4 MK at 1.3 $R_\odot$ (cf., Figure 12b). These observations preclude the existence of temperatures over 1 MK at any height near the centre of a coronal hole. Wind acceleration by temperature effects is, therefore, inadequate as an explanation of the high-speed solar wind, and it becomes essential to look for other effects, involving the momentum and the energy of Alfvén waves.

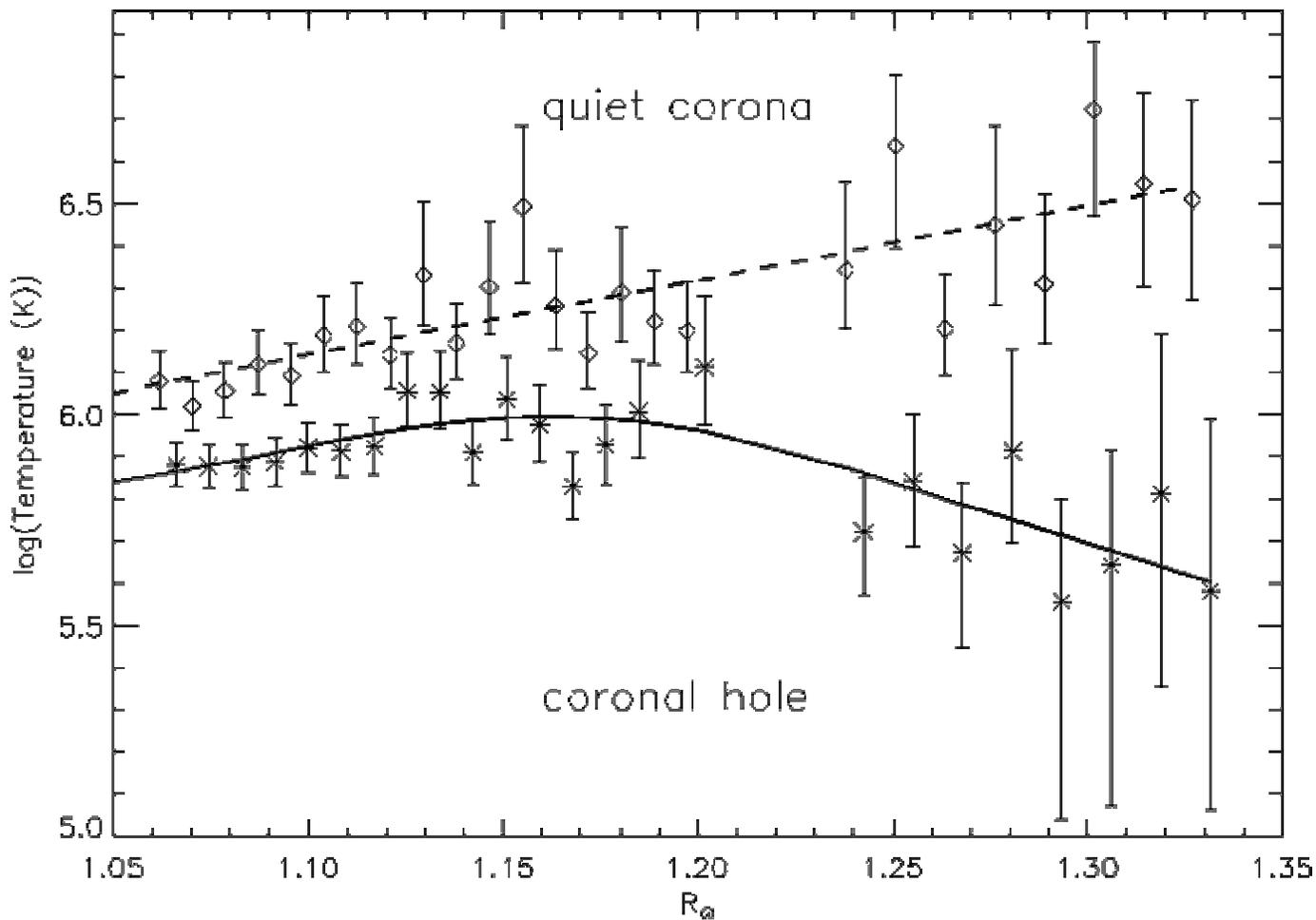

**Figure 12b.** Temperature gradient measurement in the quiet corona (equatorial west limb) and the north polar coronal hole. (From David et al. 1998).

That the solar wind is emanating from coronal holes (open magnetic field regions in the corona) has been widely accepted since the Skylab era. But there was little additional direct observational evidence to support this view. Hassler et al. (1998) found the Ne VIII emission blue shifted in the north polar coronal hole along the magnetic network boundary interfaces compared to the average quiet-Sun flow (cf., Figure 13). These Ne VIII observations reveal the first two-dimensional coronal images showing velocity structure in a coronal hole, and provide strong evidence that coronal holes are indeed the source of the fast solar wind.

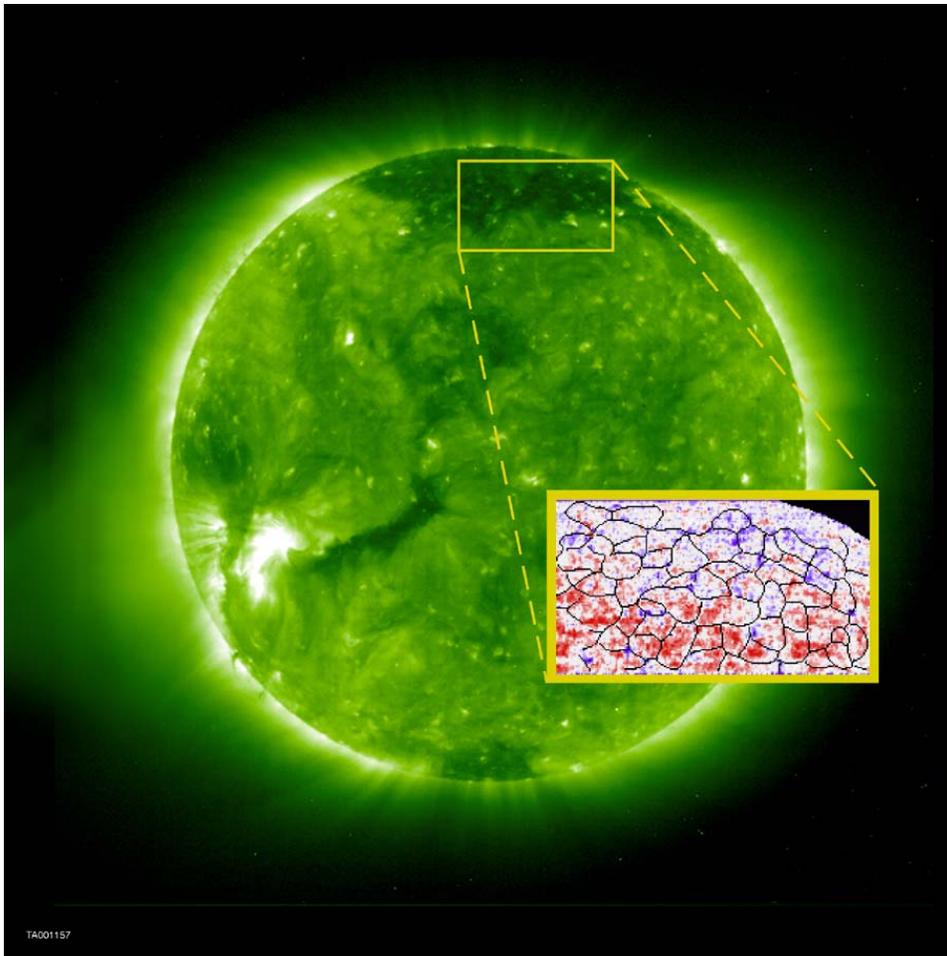

**Figure 13**. The Solar corona and polar coronal holes observed from EIT and SUMER instruments on SOHO. The "zoomed-in" or "close-up" region in the image shows a Doppler velocity map of million degree gas at the base of the corona where the solar wind originates. Blue represents blue shifts or outflows and red represents red shifts or downflows. The blue regions are inside a coronal hole or open magnetic field region, where the high-speed solar wind is accelerated. Superposed are the edges of "honeycomb"-shaped patterns of magnetic fields at the surface of the Sun, where the strongest flows (dark blue) occur. (Credit : Don Hassler and SUMER-EIT/SOHO).

Tu et al. (2005) have now successfully identified the magnetic structures in the solar corona where the fast solar wind originates. Using images and Doppler maps from the SUMER spectrograph and magnetograms from the MDI instrument on the SOHO spacecraft, they have reported the solar wind flowing from funnel-shaped magnetic fields which are anchored in the lanes of the magnetic network near the surface of the Sun (cf., Figure 14). This landmark research leads to a better understanding of the magnetic nature of the solar wind source region. The heavy ions in the coronal source regions emit radiation at certain ultraviolet wavelengths. When they flow towards Earth, the wavelengths of the ultraviolet emission become shorter which can be used to identify the beginning of the solar wind outflow.

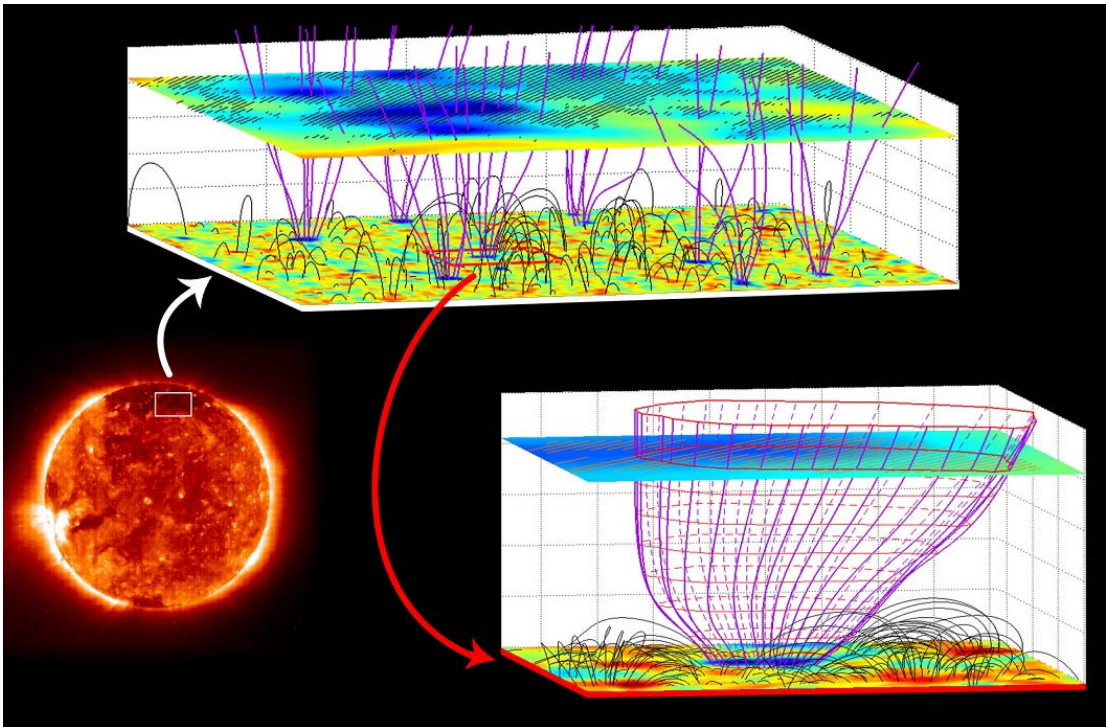

**Figure 14.** This picture was constructed from measurements which were made on September 21,1996 on SOHO with the SUMER for Doppler spectroscopy of the coronal plasma, with the MDI for magnetograms of the Sun's surface, and the EIT for the context image of the Sun.The figure illustrates location and geometry of three-dimensional magnetic field structures in the solar atmosphere. The magenta coloured curves illustrate open field lines, and the dark gray solid arches show closed ones. In the lower plane, the magnetic field vertical component obtained at the photosphere by MDI is shown. In the upper plane, inserted at 20,600 km, the Ne VIII Doppler shift is compared with the model field. The shaded area indicates where the outflow speed of highly charged neon ions is larger than 7 km/sec. The scale of the figure is significantly stretched in the vertical direction. The smaller figure in the lower right corner shows a single magnetic funnel, with the same scale in both vertical and horizontal directions. (Credit: SUMER, ESA/NASA)

Previously it was believed that the fast solar wind originates on any given open field line in the ionization layer of the hydrogen atom slightly above the photosphere. However, the low Doppler shift of an emission line from carbon ions shows that bulk outflow has not yet occurred at a height of 5,000 km. The solar wind plasma is now considered to be supplied by plasma stemming from the many small magnetic loops, with only a few thousand kilometers in height, crowding the funnel. Through magnetic reconnection plasma is fed from all the sides to the funnel, where it may be accelerated and finally form the solar wind. The fast solar wind starts to flow out from the top of funnels in coronal holes with a flow speed of about 10 km/sec. This outflow is seen as large patches in Doppler blue shift (hatched areas in the Figure 10) of a spectral line emitted by $Ne^{+7}$ ions at a temperature of 600,000 Kelvin, which can be used as a good tracer for the hot plasma flow. Through a comparison with the magnetic field, as extrapolated from the photosphere by means of the MDI magnetic data, it has been found that the blue-shift pattern of this line correlates best with the open field structures at 20,000 km.

Magnetic fields dictate the transport of charged particles. Thus solar wind particles flow along invisible magnetic field lines much like cars on a highway. When the magnetic field lines bend straight out into space (as in coronal hole regions), the solar wind acts like cars on a drag strip, racing along at high speed. When the magnetic field lines bend sharply back to the solar surface, like the pattern of iron filings around a bar magnet, the solar wind acts like cars in city traffic and emerges relatively slowly. This is well known for over thirty years and used it to give a crude estimate for the speed of the solar wind – either fast or slow. In the new work by McIntosh and Leamon (2005), the speed and composition of the solar wind emerging from a given area of the solar corona are estimated from the characteristics of the chromosphere underlying that piece of corona. Using SOHO's EIT instrument as a "finder", they isolated regions of the solar corona with open magnetic field lines (coronal holes) and closed fields (active regions). Then, using the earth-orbiting TRACE to measure the time sound waves took to travel between the heights of formation of two chromospheric continuua, they were able to demonstrate that sound travel time predicted not only solar wind speed measured by ACE but its isotopic compositon as well.

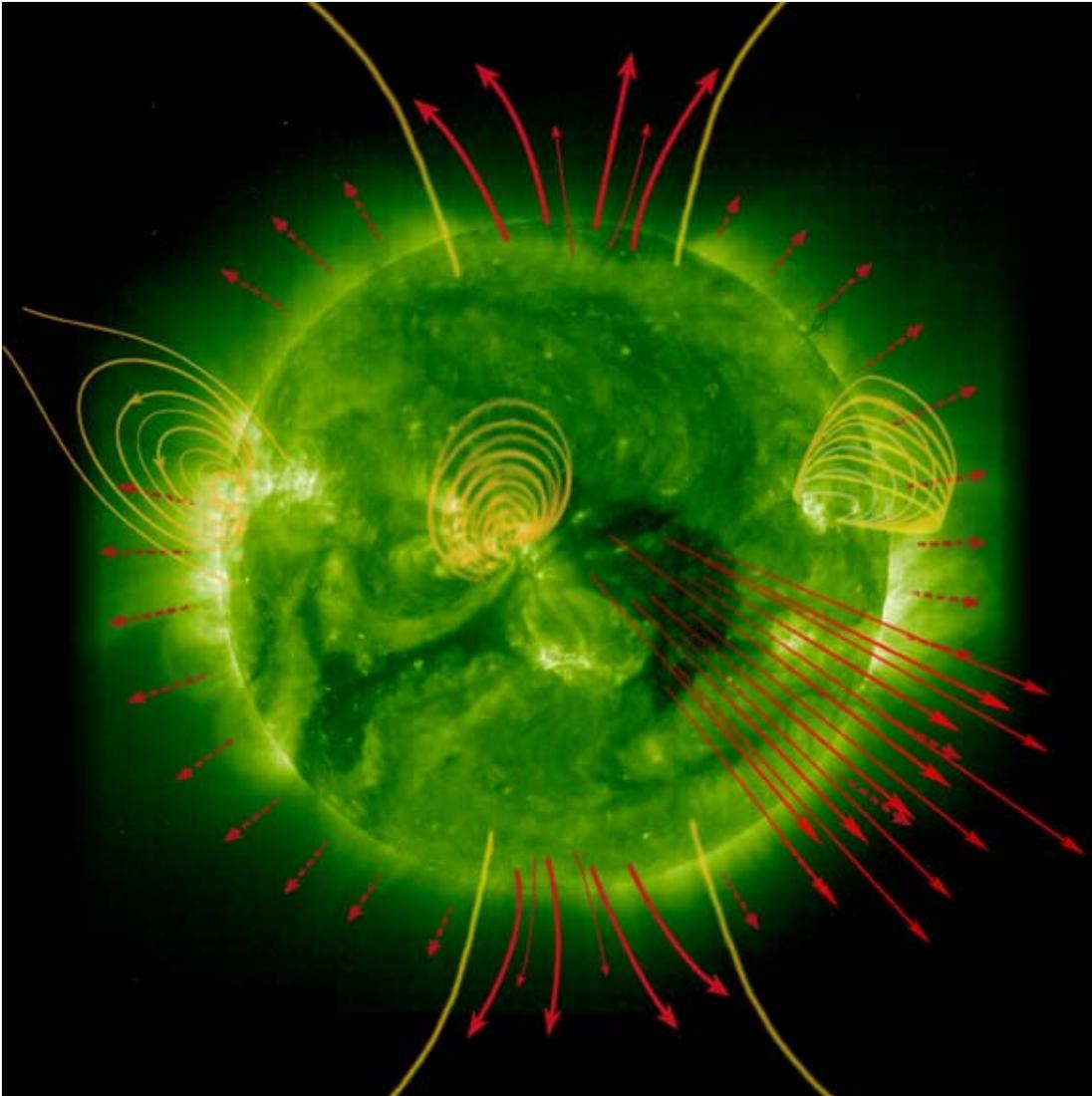

**Figure 15**. The Sun's atmosphere is threaded with magnetic fields (yellow lines). Areas with closed magnetic fields give rise to slow, dense solar wind (short, dashed, red arrows), while areas with open magnetic fields (coronal holes) yield fast, less dense solar wind streams (longer, solid, red arrows). In addition to the permanent coronal holes at the Sun's poles, coronal holes can sometimes occur closer to the Sun's equator, as shown here just right of centre. (Credit: EIT/SOHO. ESA/NASA).

The conditions in the ambient solar wind determine whether a coronal mass ejection will drive a shock wave in front of it. Shocks accelerate most of the energetic particles that can damage spacecraft and endanger spacefarers unshielded by a planetary magnetosphere. Knowledge of the state of the solar wind throughout the heliosphere is, therefore, essential to the exploration of the solar system. This work could extend solar wind predictions from the earth-Sun line (where ACE, WIND, and SOHO measure solar wind parameters) and a few planetary probes (such as the Voyagers) that also carry solar wind plasma packages, to cover the half of the heliosphere influenced by the visible hemisphere of the Sun.

## 6. Concluding Remarks

A large amount of information concerning the physical processes, taking place in the solar atmosphere has been made available in recent years from highly successful spacecraft (e.g., Yohkoh, SOHO, TRACE, and RHESSI) and from ground-based instruments. A few of the significant results from one of the SOHO's instruments, namely SUMER spectrograph, presented here alone speak of a tremendous progress made in pinpointing the processes that maintain the Sun's hot corona and accelerate the solar wind as well its source region. And the quest to unlock the Sun's mysteries goes on…

## Acknowledgements

The work is supported by the Indian Space Research Organization (ISRO) under its RESPOND programme.